\documentclass[prl,twocolumn,showpacs]{revtex4-1}
\usepackage{microtype}
\usepackage{dcolumn}
\usepackage{bm}
\usepackage{latexsym}
\usepackage{hyperref}
\usepackage{amsmath}
\usepackage[pdftex]{graphicx}

\newcommand{\mrm}{\mathrm}
\newcommand{\ket}[1]{{\left| {#1} \right\rangle}}
\newcommand{\bra}[1]{{\left\langle {#1} \right|}}

\newcommand{\vecb}[1]{\mathbf{{#1}}}

\newcommand{\Esca}{\mathcal{E}}

\newcommand{\ef}{\Esca_\mrm{eff}}

\begin{document}
\title{Mercury Monohalides: Suitability for Electron Electric Dipole Moment Searches}

\author{V.S. Prasannaa$^{1}$}

\author{A.C. Vutha$^{2}$}

\author{M. Abe$^{3,4}$}

\author{B.P. Das$^{1}$}

\affiliation{$^{1}$Indian Institute of Astrophysics, Koramangala II block, Bangalore-560034, India}
\affiliation{$^{2}$Dept.\ of Physics and Astronomy, York University, Toronto ON M3J 1P3, Canada}
\affiliation{$^{3}$Tokyo Metropolitan University, 1-1, Minami-Osawa, Hachioji-city, Tokyo 192-0397, Japan}
\affiliation{$^{4}$JST, CREST, 4-1-8 Honcho, Kawaguchi, Saitama 332-0012, Japan}

\begin{abstract}
Heavy polar diatomic molecules are the primary tools for searching for the $T$-violating permanent electric dipole moment of the electron (eEDM). Valence electrons in some molecules experience extremely large effective electric fields due to relativistic interactions. These large effective electric fields are crucial to the success of polar-molecule-based eEDM search experiments. Here we report on the results of relativistic \emph{ab initio} calculations of the effective electric fields in a series of molecules that are highly sensitive to an eEDM, the mercury monohalides (HgF, HgCl, HgBr,and HgI). We study the influence of the halide anions on $\mathcal{E}_\mathrm{eff}$, and identify HgBr and HgI as interesting candidates for future electric dipole moment search experiments.
\end{abstract}
\maketitle

Violation of time-reversal ($T$) symmetry is an essential ingredient to explain the matter-antimatter asymmetry of the universe \cite{DK03,Canetti2012}. As Standard Model sources of $T$-violation are inadequate to explain the observed asymmetry, it is imperative to look beyond it. The strongest limits on $T$-violation arising from new particles and interactions outside the Standard Model are set by searches for the permanent electric dipole moments of fundamental particles \cite{Fortson2003,Engel2013a}, like that of the electron ($d_e$). A strong constraint on the electron's electric dipole moment (eEDM), $d_e < 10^{-28} e$ cm, has been set by the experiment with ThO molecules \cite{Baron2014}, and improvements of a few orders of magnitude are forecast in the near future \cite{Tarbutt2013, Baron2014}. The eEDM experiments take advantage of the large effective electric field (often $\gtrsim 10^{10}$ V/cm) experienced by an electron in a polarized heavy polar molecule, which leads to a measurable energy shift, 
$\Delta E \propto d_e \ef$. The effective electric field, $\ef$, arises from the relativistic interactions of the eEDM with the electric fields of all the other charged particles in the molecule. This effect, whereby molecules polarized by $\sim$ kV/cm laboratory fields cause $>$ 10 GV/cm to be applied to a valence electron, is the reason for the high precision achievable in molecule-based eEDM experiments.

The value of $\ef$ for a molecule has to be obtained from relativistic many-body calculations in order to convert experimentally measured frequency shifts into eEDM values. A common heuristic that is used to estimate $\ef$ in molecules, motivated from eEDM enhancement scaling in atoms, is that $\ef \propto Z_+^3$, where $Z_+$ is the charge of the (usually heavier) cationic atom's nucleus. But molecules are not atoms. This heuristic ignores the anions which can play an important role. An improved understanding of the mechanisms leading to $\ef$ in relativistic polar molecules will lead to better choices of candidate molecules for future eEDM experiments. 

In this work, we focus on the $\ef$ for a class of heavy polar molecules, the mercury monohalides, in order to test their suitability for eEDM searches. The properties of these systems can be evaluated fairly accurately, as they have a single valence electron. The fact that they are sensitive to eEDMs in their ground electronic states (unlike molecules with metastable eEDM-sensitive states which require more complicated descriptions \cite{Fleig2014}), makes them suitable test cases for high-precision calculations. HgF has one of the largest reported $\ef$ \cite{Dmitriev1992}, making this series of HgX molecules particularly interesting as potential candidates for future eEDM experiments. The heavier Hg monohalides (HgCl, HgBr, HgI) are more electrically polarizable than HgF, which translates to a more effective use of $\ef$ and better control over systematic effects. This, in addition to better prospects for their 
production and efficient detection, makes 
the investigation of their $\ef$ values very promising for future eEDM experiments.

The expression for $\ef$ in terms of an effective eEDM operator, $H_\mrm{EDM}^\mrm{eff}$ is given by\cite{Abe2014a}
\begin{equation} \begin{split}
 \ef  = & \frac{2ic}{e} \sum_{j=1}^{N_e} \bra{\psi} \beta \gamma_5 p_j^2  \ket{\psi} \\
  = & \frac{1}{d_e}\bra{\psi} {H_\mrm{EDM}^\mrm{eff}}  \ket{\psi}
\end{split} \end{equation}

Here, c is the speed of light, e is the charge of the electron, $N_e$ refers to the number of electrons in the molecule, $\beta$ is one of the Dirac matrices, $\gamma_5$ is the product of the Dirac matrices, and $\vecb{p}_j$ is the momentum of the $j^\mrm{th}$ electron. $\psi$ is the wavefunction of a molecular state. The above expression casts the eEDM Hamiltonian in terms of one-electron operators, which makes it convenient for computations. Further details of the derivation of this form can be found in \cite{Abe2014a}. 

To obtain the molecular wavefunction $\ket{\psi}$, we use a relativistic coupled cluster (RCC) method \cite{Eliav1994,HSN2008}. The coupled cluster wavefunction can be written as
\begin{eqnarray}
\ket{\psi} = e^T \ket{\Phi_0}
\end{eqnarray}

Here, $\ket{\Phi_0}$ refers to the Dirac-Fock (DF) wavefunction, which is built from single particle four-component spinors. This is the model wavefunction for the coupled cluster calculations, and is taken to be a single determinant corresponding to an open shell doublet. $T$ is the cluster operator. In this work, we use the CCSD (Coupled Cluster Singles and Doubles) approximation, where $T = T_1 + T_2$, where $T_1$ and $T_2$ are the single (S) and double (D) excitation operators respectively. They are given by
\begin{eqnarray}
 T_1&=& \sum_{i, a} t_i^a a^\dag i \\
 T_2&=& \sum_{a > b, i > j} t_{ij}^{ab} a^\dag b^\dag j i.
\end{eqnarray}

Here, $t_i^a$ and $t_{ij}^{ab}$ are called the cluster amplitudes. In our notation, $i, j, k,\ldots$ refer to holes and $a, b, c, \ldots$ refer to particles. When $a^\dag i$ acts on a state, a hole $i$ is destroyed from that state, and a particle $a$ is created. The action of $a^\dag i$ on a model state, $\ket{\Phi_0}$ results in a state denoted by 
$\ket{\Phi_i^a}$. 

The CCSD amplitude equations are
\begin{eqnarray}
 \bra{ \Phi_{i}^{a} } e^{-T}H_Ne^{T} \ket{\Phi_0} &=& 0  \\
 \bra{ \Phi_{ij}^{ab} }  e^{-T}H_Ne^{T} \ket{\Phi_0} &=& 0
\end{eqnarray}

The term $e^{-T}H_Ne^{T}$ can be written as $\{H_N e^T\}_c$, due to the linked cluster theorem \cite{Kvasnicka1982, Bishop}. $H_N$ is the normal-ordered Hamiltonian \cite{I.Lindgren1986}. The subscript $c$ means that each term in the expression is connected. The effective fields are calculated by using only the linear terms in the coupled cluster wavefunction, since the dominant contributions come from them. Hence we evaluate
\begin{eqnarray}\label{eq:expectation_value}
 \ef &=& \langle\Phi_0\arrowvert(1+T_1+T_2)^\dag (\frac{H_\mrm{EDM}^\mrm{eff}}{d_e})_N (1 + T_1 + T_2)\arrowvert\Phi_0\rangle_c \nonumber \\
 &+& \bra{\Phi_0} \frac{H_\mrm{EDM}^\mrm{eff}}{d_e} \ket{\Phi_0}
\end{eqnarray}
We note that although the expectation value uses the linearized expansion of the coupled cluster wavefunction, the amplitudes are evaluated at the CCSD level.

We performed our calculations by combining and modifying the UTCHEM \cite{utchem,utchem2,utchem3} and the DIRAC08 \cite{dirac} codes. We used the $C_8$ point group, which reduces the computational time for the atomic-to-molecular orbital integral transformations. A summary of our calculations, both at the DF and the CCSD level, are given in Table I, and the results plotted in Figure 1. We find that the values for $\ef$ are very large for all of the chosen mercury halides, and are typically about one and a half times that of ThO\cite{Fleig2014} and about five times that of YbF\cite{Abe2014a}. This can be attributed to the fact that there is strong mixing between the valence 6s and the virtual 6p orbital.

\begin{table}[h] 
 \centering
 \begin{tabular}{|c|c|c|c|c|}
 \hline
 Molecule & Method & Basis & $T_{1,\mrm{dia}}$ & $\ef$ (GV/cm)  \\
 \hline
  HgF&DF&Hg:22s,19p,12d,9f,1g&- &104.25   \\
     &  & F:9s,4p,1d  &  &            \\ 
  HgCl&DF&Hg:22s,19p,12d,9f,1g &- &103.57   \\
      &  &Cl:12s,8p,1d &  &            \\ 
  HgBr&DF&Hg:22s,19p,12d,9f,1g &- &97.89 \\
      &  &Br:14s,11p,6d &  &           \\   
  HgI&DF&Hg:22s,19p,12d,9f,1g  &- &96.85   \\
     &  &I:8s,6p,6d   &  &            \\      
  HgF&CCSD&Hg:22s,19p,12d,9f,1g &0.0268 &115.42  \\
     &    &F:9s,4p,1d    &       &     \\ 
  HgCl&CCSD&Hg:22s,19p,12d,9f,1g &0.0239 &113.56 \\
      &    &Cl:12s,8p,1d &      &       \\
  HgBr&CCSD&Hg:22s,19p,12d,9f,1g &0.0255 &109.29 \\
      &    &Br:14s,11p,6d &       &       \\
  HgI&CCSD&Hg:22s,19p,12d,9f,1g &0.0206 &109.30  \\
     &    &I:8s,6p,6d  &       &       \\
 \hline
 \end{tabular}
 \caption{Summary of the calculated results ($\ef$) of the present work.}
\end{table}

We chose uncontracted correlation-consistent, polarized valence double zeta (ccpvdz) basis sets for F, Cl and Br \cite{bsl}, and Dyall's c2v basis sets for Hg \cite{c2v}. We use Dyall's basis for I \cite{c2v}. We use Gaussian Type Orbitals (GTO), which are kinetically balanced \cite{Dyallbook}.  Our calculations were performed without freezing any of the core orbitals. We used the following bond lengths (in nm) for our calculations: HgF (0.200686) \cite{knecht}, HgCl (0.242), HgBr (0.262), HgI (0.281) \cite{Cheung1979}.

\begin{figure}[h!]
\centering
\includegraphics[width=1.1\columnwidth]{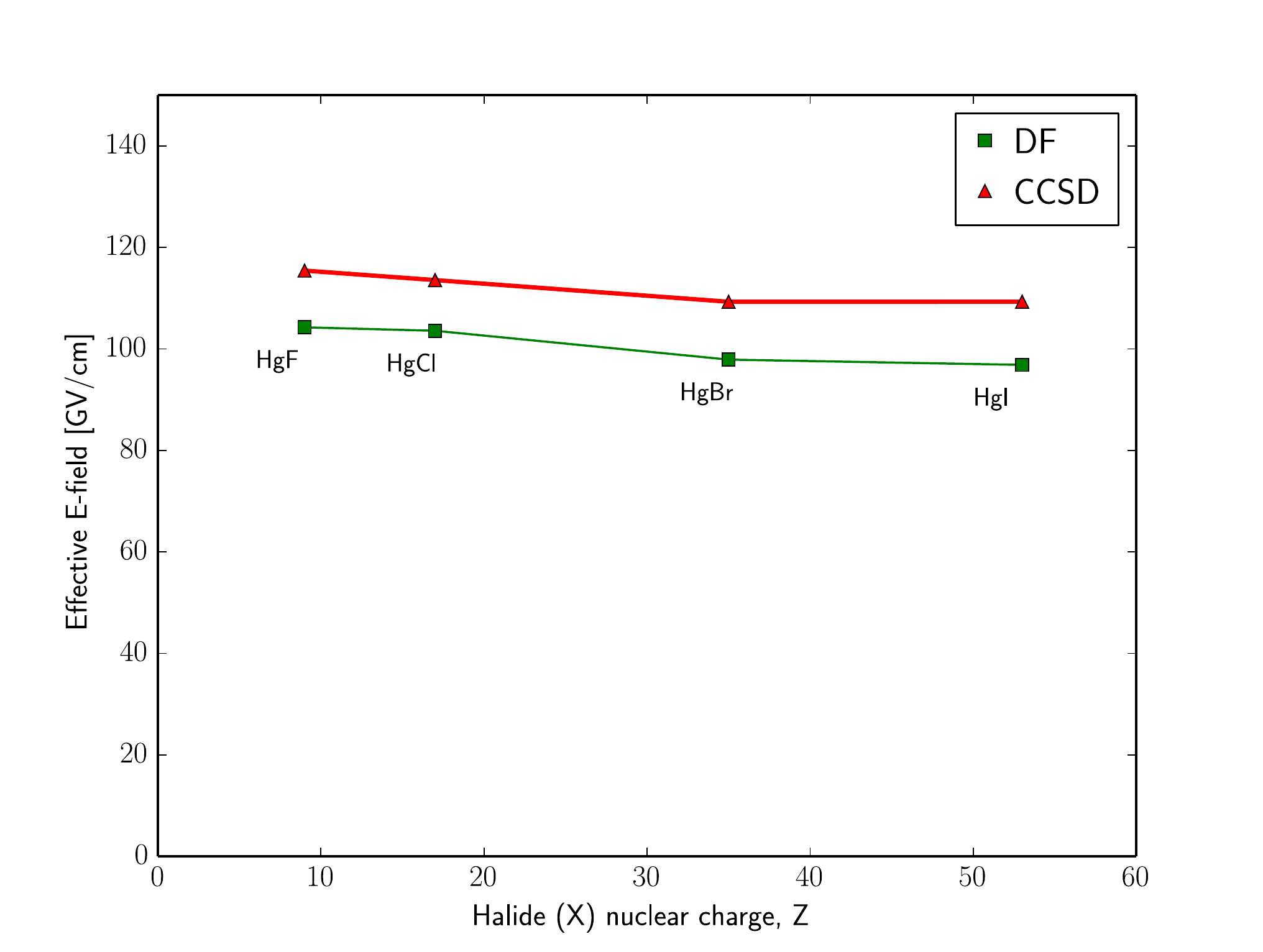}
\caption{Effective electric fields, $\ef$, for HgX molcules calculated using Dirac-Fock wavefunctions (DF, green squares), and using coupled cluster with singles and doubles (CCSD, red triangles). Solid lines are guides to the eye. The difference between DF and CCSD values indicates the contribution of electron correlation to $\ef$. }
\label{fig:Eeff}
\end{figure}

We have also reported the $T_1$ diagnostics (denoted as $T_{1,\mrm{dia}}$), a small value of which indicates the stability of single reference CCSD calculations. In the table below, we compare our result for $\ef$ in HgF with previous calculations.
 \begin{table}[h] 
 \centering
 \begin{tabular}{|l|c|}
 \hline
  Work & $\ef$ (GV/cm) \\
 \hline
  Y Y Dmitriev \emph{et al.} \cite{Dmitriev1992} & 99.26 \\
  Meyer \emph{et al.}\cite{Meyer} & 68 \\
  This work & 115.42 \\ 
 \hline
 \end{tabular}
  \caption{Effective electric field, $\ef$, in the HgF molecule.}
 \end{table}
Dmitriev \emph{et al.} computed the value of $\ef$ in HgF using relativistic effective core potential calculations. They used the minimal atomic basis set for F, and five relativistic valence orbitals $5d_{3/2}$, $5d_{1/2}$, $6s_{1/2}$, $6p_{1/2}$, and $6p_{3/2}$ for Hg. Meyer \emph{et al.} computed $\ef$ for HgF using non-relativistic software to compare their method with results obtained by other methods.

We observe from the DF and CCSD values of $\ef$ that correlation effects contribute $\sim$ 10\%. A closer analysis of the eight terms that contribute to the correlation shows that there are cancellations between some of these terms. As an illustration of this point, in Table III we show the contributions of the individual terms to the expectation value in (\ref{eq:expectation_value}) for HgF. 

\begin{table}[h] 
 \centering
 \begin{tabular}{|c|c|}
 \hline
  Term & Contribution (GV/cm) \\
 \hline
  DF & 104.25 \\
  $H_\mrm{EDM}^\mrm{eff} T_1$ &10.08 \\
  $H_\mrm{EDM}^\mrm{eff} T_2$ &0 \\
  $T_1^\dag H_\mrm{EDM}^\mrm{eff}$ &10.08 \\
  $T_1^{\dag} H_\mrm{EDM}^\mrm{eff} T_1$ &-3.91 \\
  $T_1^{\dag} H_\mrm{EDM}^\mrm{eff} T_2$ &0.22 \\
  $T_2^{\dag} H_\mrm{EDM}^\mrm{eff}$ &0 \\
  $T_2^{\dag} H_\mrm{EDM}^\mrm{eff} T_1$ &0.22 \\
  $T_2^{\dag} H_\mrm{EDM}^\mrm{eff} T_2$ &-5.52 \\  
 \hline
 \end{tabular}
 \label{tab:contributions}
 \caption{Contributions from the individual terms to the effective electric field of HgF.}
\end{table}

We see that among the correlation terms, the $H_\mrm{EDM}^\mrm{eff} T_1$ and the $T_1^\dag H_\mrm{EDM}^\mrm{eff}$ terms together contribute 20.16 GV/cm. But the $T_1^{\dag} H_\mrm{EDM}^\mrm{eff} T_1$ and the $T_2^{\dag} H_\mrm{EDM}^\mrm{eff} T_2$ terms together contribute -9.43 GV/cm. The 9 correlation terms hence add up to 11.17 GV/cm. Note that the $H_\mrm{EDM}^\mrm{eff} T_2$ term and the $T_2^{\dag} H_\mrm{EDM}^\mrm{eff}$ are zero. This follows fom the application of slater-Condon rules\cite{I.Lindgren1986} to an one-body operator, $H_\mrm{EDM}$. The same reasoning applies, for example, also for the $H_\mrm{EDM}^\mrm{eff} T_1^2$ term. The  $H_\mrm{EDM}^\mrm{eff} T_1$ term is the off-diagonal matrix element between the DF reference state, and a state with one electron excited by the electron-electron Coulomb repulsion.

The possible sources of error in our calculations stem from our choice of basis sets and not taking into account certain higher order correlation effects. 
From the difference in the effective electric field of HgF between the TZ and the DZ basis sets, we can estimate the error due to choice of 
basis to be around 1.5 percentage for all the mercury halides. A conservative estimate of the total error due to basis sets and 
omitting certain higher order correlation effects would be around 5 percentage.

\begin{figure}[h!]
\centering
\includegraphics[width=1.1\columnwidth]{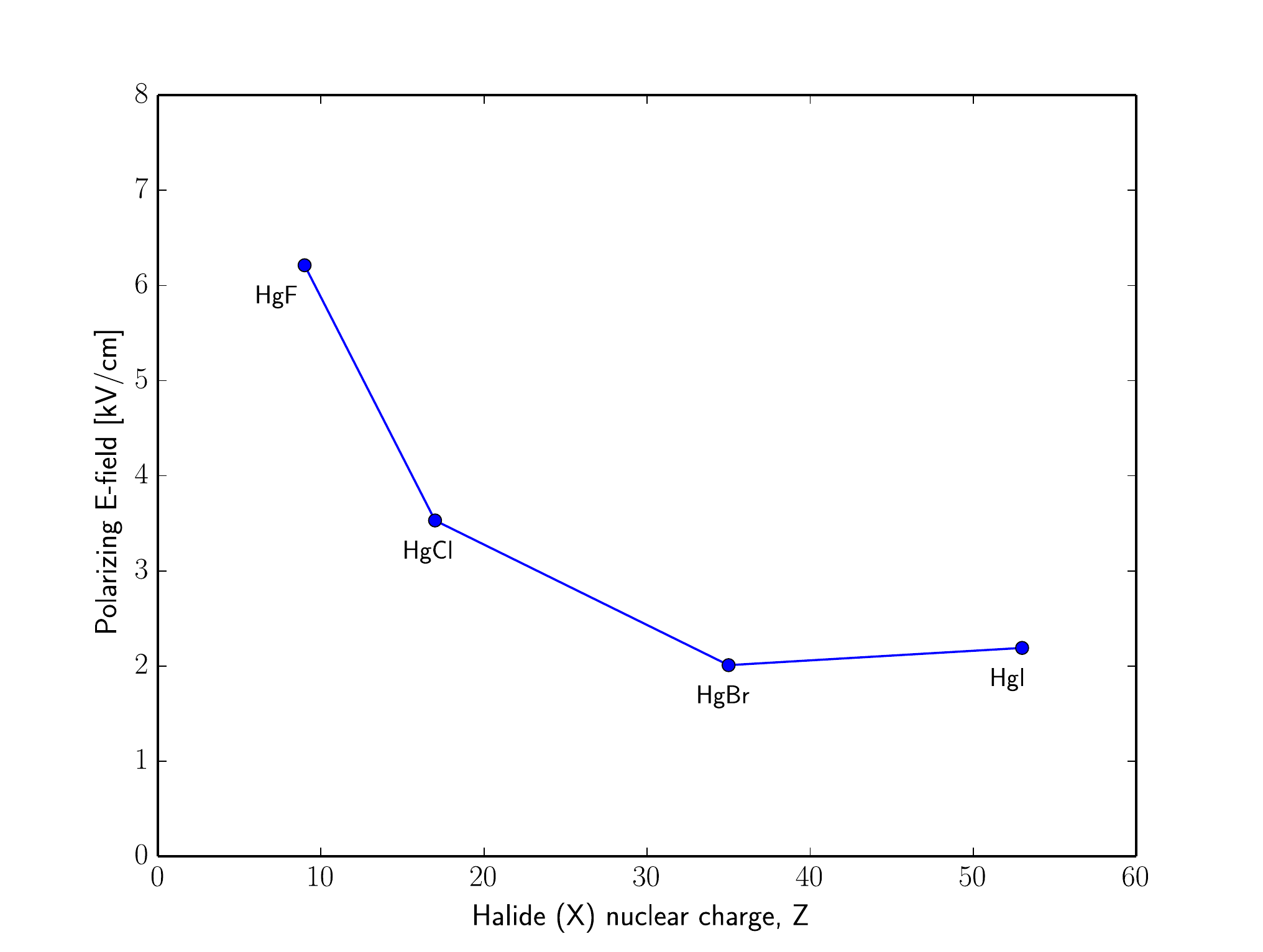}
\caption{Polarizing electric field, $\Esca_\mrm{pol} = 2 B_e/D$, for HgX molecules.}
\label{fig:polarizability}
\end{figure}

The eEDM sensitivity of experiments is $\propto \ef \sqrt{N}$, where $N$ is the number of molecules whose spin precession is detected. In addition to their large effective electric fields, there is the particularly interesting possibility that HgX molecules can be produced in large quantities at ultracold temperatures, e.g.\ by photo-association of laser-cooled Hg with magnetically trapped halogen atoms \cite{Rennick2014}. An intense and slow beam or fountain of HgX molecules could result in upto $\sim$1 s coherence time for electron spin precession. These molecules also offer a pathway for efficient detection: above their ground $X ^2\Sigma$ state, they have a repulsive $A ^2\Pi$ state which dissociates into Hg ($^1S$) and X ($^2P$) atoms. State-selective photo-dissociation of HgX, coupled with laser-induced cycling fluorescence on the product Hg atom, can be used to detect spin precession in these molecules with unit efficiency. Molecules used in eEDM experiments must be fully polarized by lab electric 
fields in order to take full advantage of their effective electric fields. The quantity that sets the scale for the required lab electric field is $\Esca_\mrm{pol} = 2 B_e/D$, where $D$ is the molecular dipole moment and $B_e$ is the rotational constant of the molecule. Figure \ref{fig:polarizability} shows the trend for HgX molecules, and picks out HgBr and HgI as attractive eEDM search candidates due to their combination of large $\ef$ and low $\Esca_\mrm{pol}$.

In summary, we have performed fully relativistic coupled cluster calculations of the effective electric fields in a family of molecules, the mercury monohalides. We find that these molecules have some of the largest effective electric fields known for polar diatomics, in addition to features that are favourable for experiments. This combination makes the mercury monohalides attractive candidates for the next generation of eEDM experiments.

\section*{Acknowledgments}
The computational results reported in this work were performed on the high performance computing facilities of IIA, Bangalore, on the Hydra and Kaspar clusters. We acknowledge Anish Parwage for his help with installing codes on the clusters. This research was supported by JST, CREST. M.A. thanks MEXT for financial support. The DiRef database was extremely useful in searching for literature \cite{diref}.

\bibliographystyle{unsrt}
\bibliography{HgX-v7}

\end{document}